\documentclass[a4paper,USenglish,11pt]{article}

\usepackage[T1]{fontenc}
\usepackage[utf8]{inputenc}
\usepackage{listings}
\usepackage{authblk}
\usepackage{url}
\usepackage{graphicx}
\usepackage{color}
\usepackage{tabularx}

\begin{document}

\title{ACGreGate: A Framework for Practical Access Control for Applications using Weakly Consistent Databases}

\author[1]{Mathias Weber}
\author[2]{Annette Bieniusa}
\affil[1]{TU Kaiserslautern, Kaiserslautern, Germany\\
  \texttt{m\_weber@cs.uni-kl.de}}
\affil[2]{TU Kaiserslautern, Kaiserslautern, Germany\\
  \texttt{bieniusa@cs.uni-kl.de}}
  \renewcommand\Affilfont{\small}

\date{}


\maketitle

\begin{abstract}
Scalable and highly available systems often require data stores that offer weaker consistency guarantees than traditional relational databases systems.
The correctness of these applications highly depends on the resilience of the application model against data inconsistencies.
In particular regarding application security, it is difficult to determine which inconsistencies can be tolerated and which might lead to security breaches.

In this paper, we discuss the problem of how to develop an access control layer for applications using weakly consistent data stores without loosing the performance benefits gained by using weaker consistency models.
We present ACGreGate, a Java framework for implementing correct access control layers for applications using weakly consistent data stores.
Under certain requirements on the data store, ACGreGate ensures that the access control layer operates correctly with respect to dynamically adaptable security policies.
We used ACGreGate to implement the access control layer of a student management system.
This case study shows that practically useful security policies can be implemented with the framework incurring little overhead.
A comparison with a setup using a centralized server shows the benefits of using ACGreGate for scalability of the service to geo-scale.
 \end{abstract}

\section{Introduction}

The ongoing globalization and digitization of services forces companies to build highly available and scalable applications that operate with low latency anywhere on earth.
By their nature, these applications need to be distributed and replicated on a global scale.
System designers choose weaker consistency guarantees for applications to gain the required performance and availability.
Typically, the data stores underlying these types of systems are key-value stores where objects are addressed by keys and have a value that can be read and updated.
To provide better performance and fail-over for possible outages, the data is replicated to different locations.
Because of weaker consistency guarantees offered by the store, operations can be issued on any of the replicas without delay and the modifications of the data state are usually asynchronously sent to the other replicas.
Data stores supporting this model are for example Amazon Dynamo\cite{decandia2007}, Riak KV\cite{riak} and Cassandra\cite{cassandra}.
But weakening consistency guarantees brings new challenges to the application design and implementation as this model makes it difficult to reason about application correctness.

\paragraph*{Access Control}

One important topic of application design is access control.
As definition of access control, we follow the definition provided in prior work\cite{eptl}.

In an application, we distinguish two kinds of operations, data operations and policy operations.
Data operations read or modify the application data, and policy operations read or modify permissions of users on objects.
Although the need for correct access control is indisputable, implementing it is a daunting task.
This is reflected by the OWASP Top 10, an index of the ten security vulnerabilities most often found in web applications.
In the OWASP\cite{owasp} Top 10 from 2013, ``Missing Function Level Access Control'' was ranked 7th, in the 2017 release candidate, ``Broken Access Control'' was ranked 4th.

Implementing access control for an application using a weakly consistent datastore is difficult because the requirements of access control seem to be in conflict with the inherent properties of weak consistency.
The programmer's intuition is that changes to permissions of a user for an object should be effective immediately in order to provide protection for subsequent data operations which may add sensitive data to the store.
Further, changing the access control rules should be possible in order to adapt the application to organizational changes.
For example, new employees may enter a company, existing employees may leave.
Even the structure of the company might change: departments might get restructured, adding or discharging responsibilities.
But despite changes, the decisions on all servers need to be consistent with the current access control rules.

For some applications, access control is deeply embedded in the design of the application.
For example, friend lists in social networks provide a mechanism to influence what data becomes visible to other users.
A classical example is the social network where Alice added her boss Bob in her friend list, and afterwards she wants to hide her photos of the last party from him.
Alice removes Bob from her friend list and afterwards uploads the photos, assuming that Bob does not have access to her photos anymore.
But on some server, the updates might get applied in a different order giving Bob temporary access to the photos.
This anomaly leaks data and is the result of the weaker ordering guarantee of the data stores.

As the examples show, the access control system should be dynamic and at the same time consistent to avoid data leakage.
These dynamic changes should also be reflected in the decisions to protect the data stored in the system from being leaked.
Operations modifying the access control policy can make up a considerable part of the overall operations performed in the system.

Weaker consistency guarantees further result in possible conflicts of concurrent permission assignments.
For correct access control, it is often not easy to see how these conflicts can be solved and which conflict-resolution strategies lead to incorrect access control decisions and hence to data leakage.

As an example, we can consider the policy that a consultant may not be responsible for two companies.
When instantiating Charly as consultant for company A, someone else might concurrently instantiate Charly as consultant for company B.
On the local server, the threat of breaking the security policy may not be detectable since the updates are initially only visible on some other replica.
When solving this concurrency conflict, one has to be very careful not to end up in a situation where Charly is consultant for both companies.

\paragraph*{Contributions}

In this paper, we discuss possible architectures to deploy access control in applications using weakly consistent data stores (Section \ref{sec:designspace}).
We present requirements for the datastore which make it possible to implement access control correctly, and we introduce ACGreGate, a Java framework for implementing such an access control layer (Section \ref{sec:implementation}).
To the best of our knowledge, ACGreGate is the first framework which allows to build access control layers for applications using a weakly consistent datastore.
Our case study shows that practical applications can be secured using ACGreGate even for complex access control policies with data dependencies (Section \ref{sec:case-study}).

\section{Access Control Layer Deployment}\label{sec:designspace}

In the introduction we have seen that many invariants that are easily implementable under strong consistency are not available in weakly consistent datastores like bounds on values due to the lack of a globally total order on operations.
How can we implement access control in environments like that?
We can either take a centralized approach where a central server is responsible for keeping the permission data consistent or we can take a distributed approach which is more complex but may yield better performance.
In the following, we discuss both approaches in detail.

\paragraph*{Central Access Control Server}

\begin{figure}
  \begin{center}
    \def\svgwidth{0.5\textwidth}
    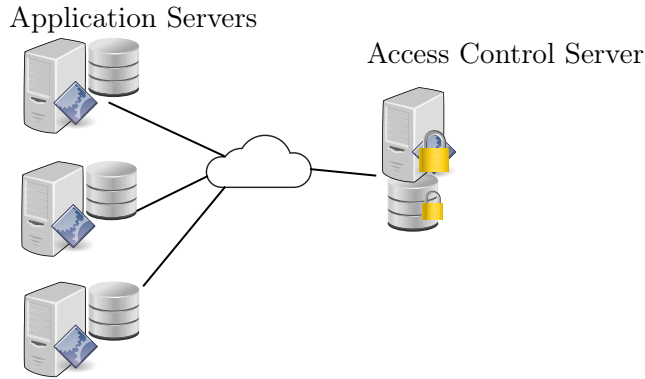
  \end{center}
  \caption{Central access control server architecture.}
  \label{fig:arch-central}
\end{figure}

Following the requirement of a consistent access control decision leads to a design as depicted in Figure \ref{fig:arch-central} where a central server makes the access control decision for all distributed copies of the application.
This architecture is common in authentication infrastructures.
The modifications of the access control rules are handled by the central access control server.
All operations performed on remote servers executing the target application are sent to the access control server to make the decision whether the operation should be accepted.
This architecture solves the problem of inconsistencies in the access control policy because there is only one copy of the policy on the central server which handles requests in a linearized way.
Since all access control decisions are made by this server, we cannot have data leakage because of inconsistent decisions.

The problem with this architecture becomes clear when  evaluating the performance (cf. Section \ref{sec:evaluation}).
Typically, the performance of an application in our scenario is bound by the performance of the storage backend.
If the access control server is not running on the same machine as this backend nor is collacated on a fast local network, the performance of the application becomes bound by the network latency.
This yields an unacceptable performance overhead compared to a system that does not implement access control.
A setup relying on a central access control server is therefore infeasible even for non-large scale systems where delays between nodes are in the order of 10 ms.

\paragraph*{Local Consistent Access Control Server}

\begin{figure}
  \def\svgwidth{0.7\textwidth}
  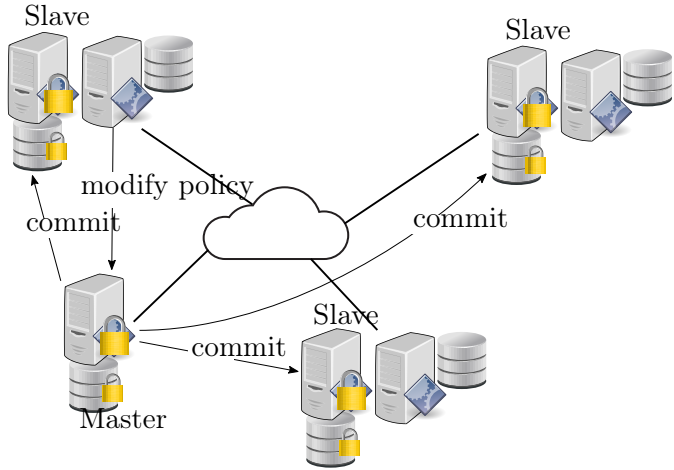
  \caption{Local consistent access control server architecture.}
  \label{fig:arch-consistentAC}
\end{figure}

What about using a primary backup model for the access control layer (cf. Figure
\ref{fig:arch-consistentAC})?
In this setting,  the policy state on all replicas is kept in sync.
But the relation between data and policy is not maintained properly.
Consider the case that we want to declassify information before publishing it.
If we apply the new policy to the old snapshot of the data, we leak classified information.
Hence, we need to keep track of all policy modifications and have to apply them consistently to their respective data snapshot.

Another problem is that the strongly consistent policy state reduces the availability.
When the application tries to modify the access control policy during a partition of the network, either the master replica or one of the slave replicas is not reachable, thus delaying the execution of the policy modification.
Since the application cannot continue without the updated policy, the application is blocked until the network partition is resolved.

\paragraph*{Combined Policy and Data State}

As the discussion shows, there is a strong inherent coupling of permissions and data.
We have already shown in prior work\cite{impl-model} that correct access control can be established by putting the policy and data state in the same weakly consistent datastore.
Our definition of access control is given in \cite{eptl} and is based on the fact that policy modifications should become active after they have been issued and until the same policy is modified again.
This property can be implemented without reverting to strong consistency or establishing global invariants.

\section{Access Control Implementation}\label{sec:implementation}

The model presented in \cite{impl-model} describes the correctness criteria for access control in weakly consistent information systems according to the above mentioned access control semantics.
But there is a considerable gap between the theoretical model and an implementation of this model.
In the following, we show how the different correctness criteria can be met by an implementation.

\subsection{Requirements derived from the Correctness Condition}\label{sec:requirements}

The theoretical model considers possible traces of the system.
Operations performed by the system need to adhere to the access control policy at the time of the operation execution.
Given a concrete trace of a system, this property is straight-forward to verify:
We iterate over the operations in the trace and compute the policy based on the policy modification performed and visible at that time and check whether the operation was permitted.

When implementing this model in a real application-level access control system, the task changes.
Instead of checking an existing trace, we need to make sure that only operations that are allowed by the current access control policy are actually executed by the system; the operations not permitted should be rejected by the access control layer.
This ensures that every resulting trace of the system is valid according to the theoretical model of correct access control.

In order to be able to implement correct access control, we will now derive several prerequisites on the context.
The correctness criterium consists of two parts that we will discuss separately: Retaining the protection relation, and correctness of the resolution of conflicts because of concurrent policy modifications.

\paragraph*{Retaining the protection relation}
All policy modifications visible when executing a data modification need to be visible on a replica, before the data operation can be applied.
The assumption is that policy modifications can protect subsequent data modifications.
By enforcing that the policy modifications is visible before applying that data modification, the model ensures that protected data remains protected and does not leak because of an outdated access control policy.
This relation between a policy modification and a subsequent data modification is known as protection relation.
As hinted in \cite{impl-model}, the protection relation is part of the per-object causality relation.
This relation holds between updates $u_1$ and $u_2$ on the same object in the data store if $u_1$ was visible when update $u_2$ was executed or the other way round.
If a data store retains the causality relation such that $u_1$ is always visible before $u_2$, we say that the data store is causally consistent.
As the example shows, support for causal consistency in the data store is one requirement for being able to implement the access control model correctly.

Causal consistency is known to be the strongest consistency criterion that can still be implemented in a highly-available way.
Several data stores implement causal consistency, for example COPS \cite{Lloyd:2011:DSE:2043556.2043593}, Orbe \cite{Du:2013:OSC:2523616.2523628}, ChainReaction \cite{Almeida:2013:CCC:2465351.2465361}, GentleRain \cite{Du:2014:GCS:2670979.2670983}, and Antidote \cite{7536539,antidote}.

\paragraph*{Conflict Resolution}
The second part of the correctness criterion for access control is concerned with solving the conflicts introduced by concurrent policy modifications in a conservative manner.
The policies are modeled in form of a set of permissions.
Each user has for each object in the store a set of permissions on this object.
These permission sets may be updated by multiple users concurrently, which can lead to inconsistencies.
The required semantics of the data type is already given in \cite{impl-model} in form of the specification of a conflict-free replicated data-type (CRDT).
For two sets of permissions $s_1$ and $s_2$ concurrently assigned for the same user and the same object, the Policy CRDT takes the intersection of $s_1$ and $s_2$.
This corresponds to taking only those permissions that both updates agreed on.
This semantics can be easily implemented in a data store supporting CRDTs by taking the implementation of a multi-value register and modifying the read operation.
A multi-value register retains all concurrently assigned values.
Assignments to the register only replace those values visible when the operation is executed on the data store.
In case of the Policy CRDT, this yields both sets of permissions that have been assigned concurrently.
We can compute the intersection of these sets as the result of the read operation.

\subsection{Requirements of an Online Check}


In the theoretical model, the decision procedure operates on the same snapshot of data and policies as the operation to be executed.
When implementing the model, we have to make sure to preserve this relation between policy and data state.
Separating the operations and thereby working on different snapshots of data and policy state can lead to incorrect behavior.

One scenario is that we base our decision on an outdated policy.
As already discussed with relation to the theoretical model, policy modifications are usually used to protect data in the store.
A policy modification might restrict access to confidential information which was added to the store after the policy modification.
If this restriction is not reflected in the \emph{outdated policy}, the access to the confidential information might be granted to an unauthorized user.
The other way round, a data modification might have declassified information which is then safe to be read.
If we operate on \emph{outdated data}, this data might still reflect the confidential information because the declassifying operations are missing, which again leads to data leakage.
Other replicas accept concurrent operations and these operations can in general get visible at the local replica at any time.
The conclusion is that there is no safe order in which we can read the policy and execute the operation to be checked.
Both operations need to happen atomically.

Atomic operations are supported in some weakly consistent data stores as highly available transactions (HATs) \cite{burckhardt_eventually_2012-1,bailis_highly_2013-2,7536539}.
HATs support atomicity without reducing availability.
The main difference to strongly consistent transactions is that HATs do not guarantee serializability of the updates, but the usage of CRDTs solves the problem of conflicting concurrent updates.
The Cure protocol\cite{7536539} for example supports transactional causal+ consistency, an extension of causal consistency with highly available transactions.

To summarize, in order to implement our framework for correct access control in applications using weakly consistent data store, we require from the underlying datastore to support causal consistency, CRDTs as data model and transaction support for atomic multi-object updates.
All of these properties are supported by the Cure protocol\cite{7536539}.

\subsection{Framework Implementation}

\begin{figure}
  \def\svgwidth{0.9\textwidth}
  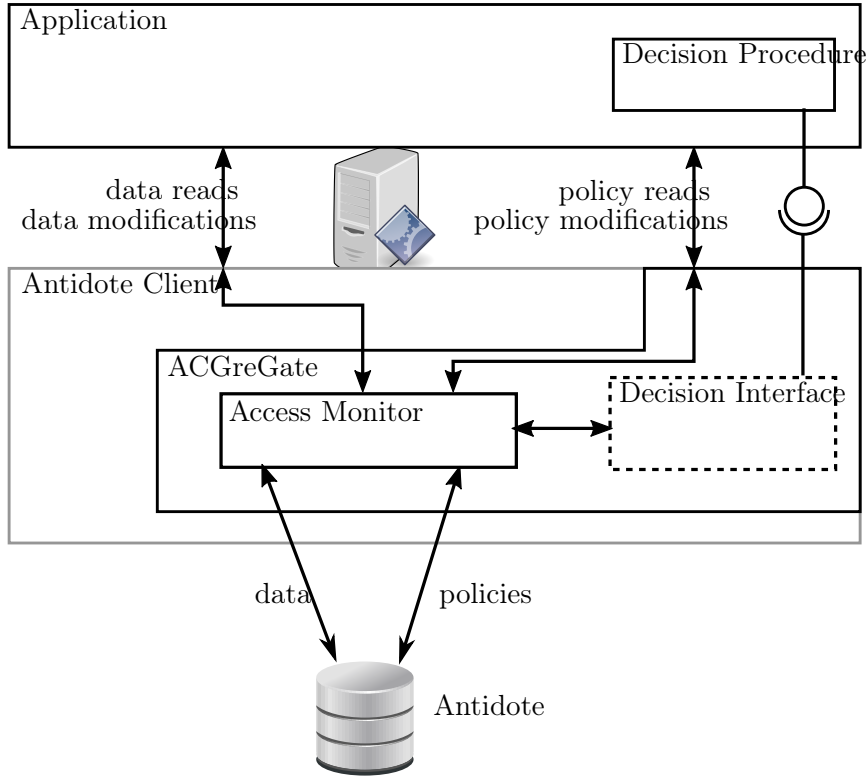
  \caption{Architecture of ACGreGate.}
  \label{fig:arch-acgregate}
\end{figure}

ACGreGate is a framework for correct access control layers of applications written in Java, that use Antidote as datastore.
Antidote\cite{antidote} is an open-source implementation of the Cure protocol written in Erlang with a variety of client libraries for different programming languages.
We built our framework ACGreGate on top of this provided infrastructure for evaluation.

We implemented the Policy CRDT (cf. Section \ref{sec:requirements}) and included it in Antidote's CRDT library.
Antidote offers a Java client, which allows applications to connect to Antidote and execute update and read operations based on the provided CRDTs.
We implemented ACGreGate as an extension to this Antidote Java client library.
The architecture of ACGreGate (Figure \ref{fig:arch-acgregate}) is described in the following paragraphs in detail.

\paragraph*{Decision Procedure}

To implement the online check, ACGreGate intercepts all operations sent by the client library to the Antidote database and processes them by an access control monitor.
This monitor takes an implementation of a decision procedure to make the access control decisions.
The decision procedure is application dependent and corresponds to the security policy of the application.
In principle, the decision procedure takes the operation, the currently acting user, and the permissions of this user and decides, whether the operation is allowed to be executed or not.
In the framework implementation, the decision procedure is implemented as a class implementing the \lstinline!DecisionProcedure! interface which can be seen in Listing \ref{code:decision-procedure}.

\begin{lstlisting}[caption={The DecisionProcedure interface.},label=code:decision-procedure,float,language=Java]
public interface DecisionProcedure {
  boolean decideRead(ByteString currentUser, AntidotePB.ApbBoundObject object, Object userData, SecurityLayers layers);

  boolean decideUpdate(ByteString currentUser, AntidotePB.ApbBoundObject object, AntidotePB.ApbUpdateOperation op, Object userData, SecurityLayers layers);

  boolean decidePolicyRead(ByteString currentUser, AntidotePB.ApbBoundObject key, ByteString user, Object userData, SecurityLayers layers);

  boolean decidePolicyAssign(ByteString currentUser, AntidotePB.ApbBoundObject key, ByteString user, Collection<ByteString> newPolicy, Collection<ByteString> oldPolicy, Object userData, SecurityLayers layers);

  LayerDefinition requestedPolicies(ByteString currentUser, AntidotePB.ApbBoundObject object);
}
\end{lstlisting}

The decision procedure is split up for convenience into methods to decide about read and update operations as well as policy read and policy assign operations.
Depending on the type of operation, different parameters are provided to the decision procedure, most notably the user performing the operation and information about the operation itself.
Access to the current permissions of the acting user is provided using the concept of \emph{security layers}.
The concept behind security layers is best described based on an example.

In our model, each user has a single set of permissions on an object.
In practice, many applications have hierarchical data structures and users are given permissions on different layers of this hierarchy.
A simple example is a university with lecturers.
A lecturer gives a lecture and has full access to all data regarding this lecture.
A teaching assistant supervises the exercise of the lecture. Tutors may be employed to mentor a group of students in a particular exercise session.
Regarding the data of a student participant, users may have permissions from different levels.
A lecturer has the permission to access the data of any participant of the exercise if she reads the corresponding lecture.
An assistant of the exercise has permission to access the participant's data if he is the assistant of the participant's exercise.
Finally, a tutor may access the participant's data only if she is the participant's tutor.
In the datastore, this hierarchy can be split into objects with key references.
If we would only allow a single permission set per user and object, we would have to copy all permissions on the lecture to the exercise of this lecture and copy all permissions of the exercise to all participants of this exercise.
For registering new participants, we would have to set the permissions on the lecture and exercise, as well.
This approach is error prone and complex.
Instead, we allow to define the hierarchy for an object in form of a \lstinline!LayerDefinition!.

The definition of a layer consists of a name for the layer and the key of the object the layer refers to.
A \lstinline!LayerDefinition! can contain multiple layers.
The definition of the layers is transformed into a \lstinline!SecurityLayers! object which allows access to the permissions of the currently acting user on all objects defined in the the \lstinline!LayerDefinition!.
The complete set of permissions for a user on an object can be computed by taking the union of the permissions on all layers.
This approach allows to avoid redundancies in the permission assignments, but still gives access to the complete set of permissions in a well-defined way.

\paragraph*{Operation Constraints}

The key-value data stores we consider support CRDTs which are in essence data types that can safely be used in an unsynchronized distributed manner.
The operations on a CRDT object are typically on a higher level than in strongly consistent datastores.
For example, Counter CRDTs allow to increment or decrement the counter and to read the current value, Set CRDTs allow to add and remove elements and Map CRDTs allow to update the value of a specific key in the map.
Access control policies on the data level of such a datastore usually restrict these operations and describe which properties the operations should have.

For maps, we might want to restrict which keys should be updatable and which key bindings should be removable.
For sets, we might want to restrict which elements should be addable or removable.
ACGreGate supports implementing these constraints in the implementation of a decision procedure by providing a domain specific language to describe these restrictions.
An example of such constraints can be seen in Listing \ref{code:constraint-language}.

\begin{lstlisting}[caption={Constraint language examples.},label=code:constraint-language,float,language=Java]
AntidotePB.ApbUpdateOperation op = ...;
String studentid = ...;

boolean mayParticipate = isMapUpdate(and(
  assignsOnly("participants"),
  constrainAssigns(keyConstrain("participants",
    isSetUpdate(
      or(
        and(
          setAddsOnly(studentid),
          noSetRemoves
        ),
        and(
          setRemovesOnly(studentid),
          noSetAdds
        )
      )))),
    noMapRemoves)).appliesTo(op);
\end{lstlisting}

The constraint specifies that the operation needs to be a map update (\lstinline!isMapUpdate(...)!).
The additional constraints are that only the key \lstinline{"participants"} may be updated and no key-value mappings may be removed (\lstinline!noMapRemoves!).
The update of the key \lstinline{"participants"} is again restricted to be a set update (\lstinline!isSetUpdate(...)!).
Additional constraints regarding the update of the participant set are that either the student identification number may be added to the set (\lstinline!setAddsOnly(studentid)!) and no elements may be removed from the set (\lstinline!noSetRemoves!), or the other way round.
This constraint can be checked against the operation provided as parameter to the decision procedure (\lstinline!.appliesTo(op)!).
In the context of the case study presented in Section \ref{sec:case-study}, the semantics of the constraint is that students can add themselves as participants to an exercise group.

\paragraph*{Datastore Layout}

To guarantee atomicity and causal consistency of data and policy updates, we persist the permission sets together with the data in the datastore.
Because the permissions should only be modifiable using special access control operations, these sets need to be isolated from the other data in the data store.
Antidote supports buckets as name spaces for keys in the key value store.
These buckets are used in ACGreGate to achieve the isolation between policy and data state.
The bucket of the intercepted operations are prefixed with distinct prefixes depending on wether the operation is a data or a security operation.
The permission sets are saved per object and user by generating a key based on the object key and the user identifier.
This approach guarantees complete isolation between policy and data state and avoids accidental or malicious modification of the permission sets.

\paragraph*{Interface Modifications}

To allow easy replacement of the standard client library with ACGreGate, we kept the interface as close as possible to the original client interface.
Only minor modifications with respect to the transaction interface are needed in order to specify the currently acting user.
When starting a new transaction, the identifier representing the current user has to be passed as a parameter.
All operations executed in this transaction are executed under the name of this user.
All other modifications are done by returning subtypes of the original client library types such that ACGreGate can act as a substitute to the original client library.
The type \lstinline!AntidoteClient! needs to be substituted by \lstinline!SecureAntidoteClient!, the type \lstinline!Bucket! needs to replaced with \lstinline!SecuredBucket!.

\paragraph*{Data Access}

In practice, access control policies may depend on the data state of the application.
A concrete example of an access control policy that needs this feature is self-registration for an exercise in a lecture management system.
Students are allowed to register themselves for an exercise while the exercise registration is open.
The flag that signals that the registration is currently open is usually part of the data state of the application, yet it has influence on the access control decision.
In this case, the decision procedure implementation will access the object value of the exercise layer and check the status flag for the exercise registration.

This feature enables additional and powerful policies which enable to provide also ownership and attribute-based access control.
To implement an ownership model of objects in the database, one could associate a register-type attribute with each object that holds the current owner of the object.
The decision procedure can access this attribute and decide on the concrete permissions.

ACGreGate supports policies involving data state by giving the decision procedure access to the current value of a layer object.
However, policies involving data state are not covered by our theoretical model in \cite{impl-model} and as such, there is no guarantee that enforcing these policies avoids data leakage in all cases.
Instead, the safety of the policy with respect to data leakage has to be reasoned about on a per-application and policy level.

\section{Case Study}\label{sec:case-study}

We show the applicability of ACGreGate to practical access control policies by implementing the access control layer of a student achievement tracking system (STATS).
In the following, we describe the data structure and the access control policy of STATS.

\subsection{Data Model}

The purpose of STATS is to manage the grading of exercises associated to lectures together with the corresponding exams.
A student can register an account which includes personal data such as the name, the student identification number, and an email address.
At the begin of each semester, students can register for participation in an exercise.
The status of the exercise registration is indicated by a flag in the exercise object.
The students registered for an exercise are organized into groups for which a time slot on a specific day is allocated on which a class is given in a specified location.
One or more assistants are responsible for the exercise organization.
The exercise classes are given by tutors which are responsible for several exercise groups.
Students can obtain points for homework submissions which are persisted in form of a map from sheet id to the achieved points.

Similarly, students can register and participate in exams.
An exam can be an open exam, for example a trial exam, which is indicated by a flag, or it is only open for students who participated successfully in the exercises.
In the end, the exam results together with the grade assignment are published which is indicated again by a flag in the corresponding exam object.

\subsection{Access Control Policy}

The STAT System was developed in the context of  maintaining the integrity of structured data \cite{michel_formal_2014}.
The access control policy of the STAT System was formulated in terms of integrity constraints. Therefore, only updates were originally restricted.

The access control policies for the update operations are defined as follows.
Admins are allowed to perform all updates.
Assistants of an exercise are allowed to modify the attributes of the exercise as well as open and close the exercise registration.
They are also allowed to create and update groups and sheets of the exercise as well as delete them.
Additionally, assistants are allowed to assign students to groups and take them out again as well as to assign and remove tutors.
Tutors are allowed to assign students of their group to teams and to modify the results of students in their group.
Examiners can modify the attributes of the exam and can add and remove participants.
They can create, modify and delete tasks and grades and assign results to participants.
Examiners can also open the exam for registration for students and publish the results of the exam for the participants.
Students can always register for an exercise, but they can only sign-up for or sign-out of a group if the exercise registration is open.
Students can also register for and unregister from exams that are open to self-registration.
We now added further the access control rule that assistants and tutors can see the results of their groups and that examiners can only see the results of their own exams.
Students have only access to their personal results of the exercises and exams.

These policy rules are based on the application functions.
To be implemented as a decision procedure in ACGreGate, they have to be translated to rules on the data update level.
For example, an update of an exercise object is allowed if the user is an admin, an assistant of this exercise or a student such that the exercise adds the student id to the set of registered students for the exercise.
We did this for all rules stated above and implemented a decision procedure according to these rules.
The result is a Java method with less then 250 lines of Java including error handling and logging.
Even though the rules include dependencies to data values, the system is secure in the sense that it avoids data leakage:
\begin{itemize}
\item The exercise and exam registration flags control, whether students can register themselves; these flags do not expose additional data.
\item Setting the publish flag in the exam exposes the exam results to the students.
Assigning this flag to different values concurrently might expose the exam results without consensus. However, we argue that this situation does not arise in practice. Typically, a single examiner sets the flag on behalf of all examiners once they decided to publish the results.
\end{itemize}

\section{Evaluation and Results}\label{sec:evaluation}

The STAT System has been in active use in the CS department at the University of Kaiserslautern in Germany since 2011, though the current implementation is based on a different data persistence layer.
We captured the data state of roughly two years with four exercise iterations and six exams.
A workload generator takes this data and recreates the same data state in the Antidote backend of STATS.
Since the application performs consistency checks, this workload triggers 102\,861 datastore read operations and 32\,991 datastore update operations.
The execution of the decision procedure triggers additional read operations for retrieving the current policy and additional data for some rules.
This yields additional 216\,923 read operations, which is an overhead of about 135\% over the actual operation.
The tests we report on here were performed on a server with two Intel Xeon E5-2620 CPUs and 64\,GB of RAM running Ubuntu 16.04.1 LTS. We used Docker to build images of the Antidote database and the STAT System. The Docker version used was 1.12.6.

We performed two types of tests.
In the first setup, we started one container with the STAT System with the access control layer implemented using ACGreGate.
This container was connected to a container running Antidote using a Docker virtual network.
The second setup used a modified version of the STAT System that forwarded the access control decisions to an access control server implementing the STATS policy running as a separate container.
The STATS container was connected to the Antidote container using one virtual network and connected to the access control server using another separate virtual network.
On this network, we used the netem network emulator to simulate different latencies between the application and the access control server.

\begin{table}
  \caption{Performance results.}
  \label{tab:performance}
  \begin{tabularx}{\textwidth}{|X|X|X|X|X|}
    & \multicolumn{1}{c|}{\bfseries net delay} & \multicolumn{1}{c|}{\bfseries request delay} & \multicolumn{1}{c|}{\bfseries throughput} & \multicolumn{1}{c|}{\bfseries duration} \\ \hline
    ACGreGate & -       & -         & 1\,079.2\,ops/s & 2:06 \\
    nodelay   & 0\,ms   & 0.3\,ms   & 1\,257.9\,ops/s & 1:48 \\
    small     & 10\,ms  & 13.9\,ms  &     75.9\,ops/s & 29:51 \\
    medium    & 50\,ms  & 69.0\,ms  &     16.3\,ops/s & 2:19:11 \\
    large     & 100\,ms & 136.8\,ms &      8.3\,ops/s & 4:34:21 \\
  \end{tabularx}
\end{table}

The results of these experiments are given in Table \ref{tab:performance}.
The net delay refers to the emulated delay of the network between the application and the access control server.
The throughput is calculated for the operations issued by the application excluding the operations produced by the decision procedure.
The throughput in the ACGreGate and nodelay tests are bounded only by the performance of a single Antidote node.
When adding network delay, this delay dominates the performance.
Even with only 10\,ms roundtrip delay, the performance drops from almost 1\,300 operations per second to about 76 operations per second.
This delay is roughly observed for servers in different nearby cities in the same country.
For more than 50\,ms of roundtrip delay on the network, the performance drops by two orders of magnitude to about 16 operations per second, which makes using a central access control server infeasible for cross-country or even geo-scale replication.

\begin{figure}
  \includegraphics[width=\textwidth]{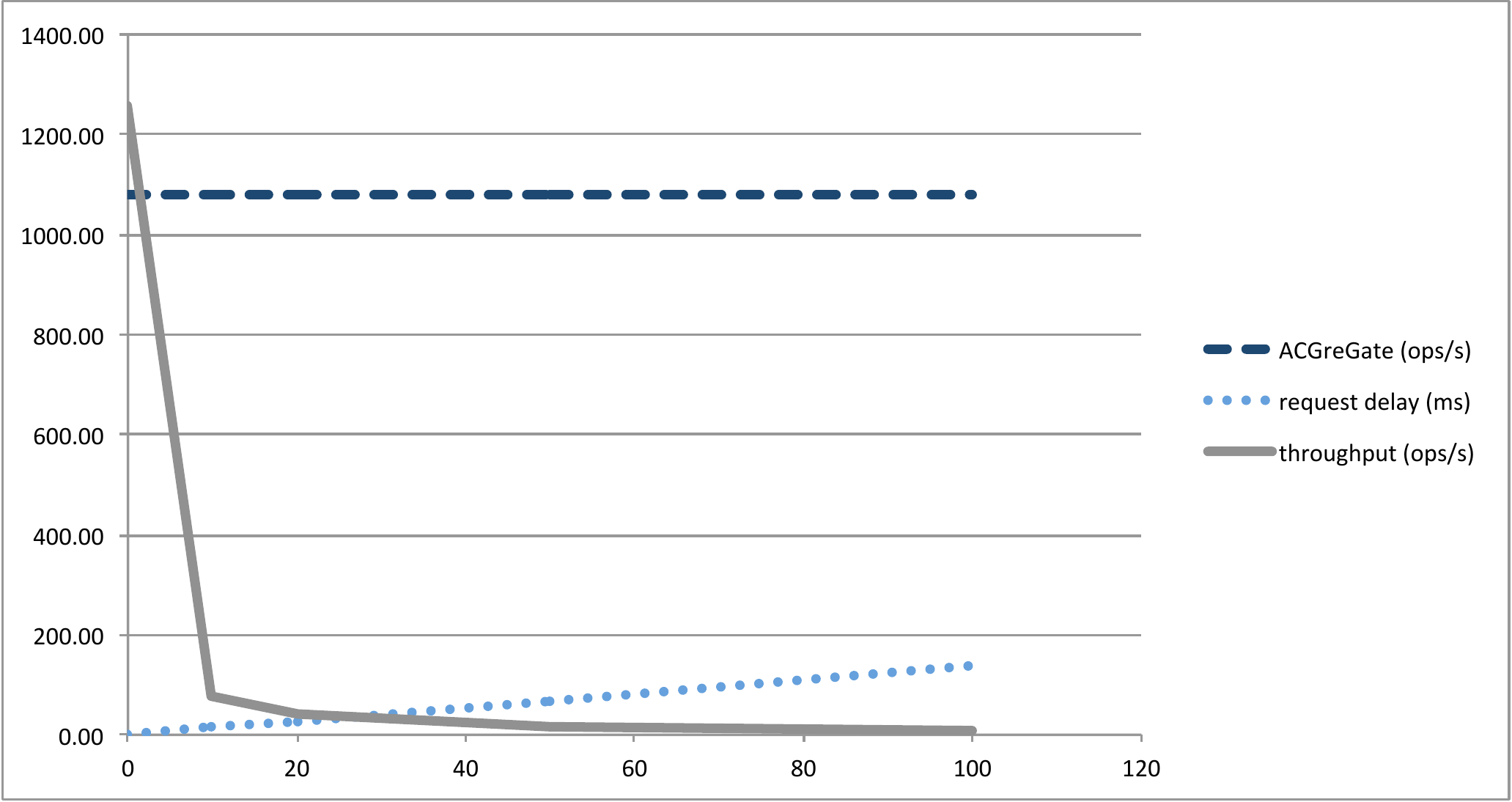}
  \caption{Performance comparison for different network delays.}
  \label{fig:performance}
\end{figure}

The graph in Figure \ref{fig:performance} shows a comparison between the performance on ACGreGate and the performance of using a central access control server.
ACGreGate has direct access to the policy and gets an immediate response for reading the permissions from the local state in the Antidote datastore.
This makes the performance independent of the network delay.
In contrast, the centralized architecture is very sensitive to network delay.
For a network delay of 0\,ms, the centralized architecture performs slightly better then ACGreGate in our test.
The reason for this is that we used a very fast in-memory data store to implement the access control server whereas ACGreGate reads the policies from Antidote.
Using a persistent database, for example Antidote, to implement the access control server would yield a result equivalent to the performance of ACGreGate for the 0 ms delay case.

When running the workload generator without access control checks enabled, the throughput is at roughly 1750\,ops/s and the program runs in 1:17 minutes, which shows a slowdown for ACGreGate of about 40\,\%.
This slowdown is not unexpected and depends on the number of requests for additional information required by the decision procedure.
The performance could further be improved by co-locating data and security attributes or caching of access control decisions.

There are access control architectures that influence the performance far less but also offer less security guarantees.
A role-based access control system can be setup with a central authentication server that also manages the roles of users.
When authenticating to the server, the user opens a session for which the roles of the user are active that were valid when performing the authentication.
This setup can make access control decisions locally without contacting the server again and without reading additional data from the store.
But modifications of user permissions do not become effective together with the data modifications, but only after the next authentication request of the user.
In addition, role-based access control is far more restrictive with respect to the security policies than the model used by ACGreGate.

\section{Related Work}

\paragraph*{Access Control for Applications using Weakly Consistent Data Stores}

The topic of access control for applications using weakly consistent data stores has received surprisingly little attention.
The original version of Amazon Dynamo \cite{decandia2007} did not offer authentication and authorization capabilities.
Several other related eventually consistent data stores offer meanwhile techniques to implement access control, but the granularity is not fine enough to provide access control on the application level.
Riak KV \cite{riak}, MongoDB \cite{mongodb}, Couchbase \cite{couchbase} and Cassandra \cite{cassandra} all support the management of users, roles and permissions.
But the smallest granularity is on the level of buckets or collections, comparable with tables in relational data bases.
Typical permissions on this level allow to read, write, modify, or delete any value of the bucket or collection.
A more fine-grained permission level relating to the operations on the application level is not supported.
MongoDB Stich \cite{mongodbstitch} is a framework for applications built on MongoDB that provides support for access control.
The policies supported are mainly based on the current data state and correctness of the access control system is not clear for any definition of access control.

Samarati et. al\cite{samarati1996} describe a high-level approach to authorization in eventually consistent systems.
The general idea is to optimistically accept all operations and compensate the operations which were executed despite the security policy by performing rollbacks.
While this approach guarantees convergence of the security policy, it is not clear for each operation how to undo the effect of this operation after it has been executed.
One of the problems is the potential binding between operations and effects in the data store and changes of the real world.
For example, a banking system allows to withdraw money from an account and the ATM outputs the money.
In this case, it is hard to undo the withdrawal because the person with the money has already walked away.
In addition, the guarantees given by such an optimistic system remain unclear.
Effects of operations can be perceived by a user of the data store before the rollback, thereby possibly leaking sensible information.

Wobber et. al\cite{wobber2010} present an access control model for weakly consistent, mutually distrustful replicated systems.
Their focus of work is on partial replication with different access policies per replica.
While we consider a different setting of fully replicated systems, similar problems can be identified.
The causality between a policy and the subsequent operation that are permitted by the policy is captured in their model by waiting for the required policy change to arrive.
However, the causality between a policy change that restricts the visibility of the effect of an operation and the subsequent execution of this operation is not captured.
As such, the model still allows leaking sensitive information because of the possible violation of the protection relation between a policy change and a subsequent data operation.

\paragraph*{Access Control for Distributed Applications}

In the area of access control for distributed applications, most of the work applies to strongly consistent systems.
Bui et. al\cite{bui_fast_2017} developed an algorithm named FACADE for fast evaluation of stateful attribute-based access control policies.
The policies supported by the algorithm are more expressive than the policies we consider.
In their work, access control decisions can be influenced by prior access control decisions which requires a linear order of operations and additional state to coordinate the distributed access control decisions.
The performance impact of the evaluation is too large for the high-performance applications we consider, the author give an average latency of about 37\,ms for their algorithm.
In addition to that, the coordination needed to support stateful access control policies reduces the availability of the application as described in Section \ref{sec:designspace}.


\subsection{Conclusion}

Providing correct, low-latency access control is a challenge for developers of highly available and scalable systems.
A careful analysis of the intricate interplay of data and policies shows that such an access control system can be based on causally consistent data stores with support for highly available transactions and conflict resolution of concurrent updates.
To show the feasibility of such a system, we implemented ACGreGate, a Java framework for applications based on the Antidote data store.
Our case study shows that ACGreGate allows to implement non-trivial policies encountered in systems in active use today.
As the use case shows, using ACGreGate to implement access control in an application limits the scalability and performance only to the scalability and performance of the datastore.



\bibliography{references}

\end{document}